\documentclass{KapProc} % Computer Modern font calls

\usepackage{epsf}
\usepackage{graphicx}
\normallatexbib

\begin{document}

\articletitle{Theory of STM Spectroscopy of\\ Kondo Ions
on Metal Surfaces}

%\articlesubtitle{This is an Article Subtitle}

\author{O. \'Ujs\'aghy$^{a,b}$, J. Kroha$^c$, L. Szunyogh$^a$ 
and A. Zawadowski$^{a,b,d}$}
\affil{
$^a$Department of Theoretical Physics and $^b$Research Group of 
the Hungarian Academy of \hspace*{0.3cm}Sciences, 
Technical University of Budapest,
H-1521 Budapest, Hungary\\
$^c$Institut f\"ur Theorie der Kondensierten Materie, Universit\"at 
Karlsruhe,\\ \hspace*{0.3cm}D-76128 Karlsruhe, Germany\\
$^d$Solid State and Optical Research Institute of the Hungarian
Academy of Sciences,\\ \hspace*{0.3cm}H-1525 Budapest, Hungary}
\email{ujsaghy@born.phy.bme.hu}

\chaptitlerunninghead{Theory of STM Spectroscopy of Kondo ions  \dots}
\begin{keywords}
Kondo effect, Fano resonance, scanning tunneling microscopy
\end{keywords}

\begin{abstract}
The conduction electron density of states nearby a single magnetic impurity,
as measured recently by scanning tunneling microscopy (STM),
is calculated. 
It is shown that the Kondo effect induces a narrow Fano resonance 
as an intrinsic feature in the conduction electron density of states. 
The line shape varies with the distance between STM tip
and impurity, in qualitative agreement with experiments, and
is sensitive to details of the band structure.
For a Co impurity the experimentally observed width and shift of the
Kondo resonance are in accordance with those obtained from a combination of
band structure and strongly correlated calculations.
\end{abstract}

\begin{figure}[ht]
\centerline{
\epsfxsize=5.5cm
\epsfysize=3.5cm
\epsfbox{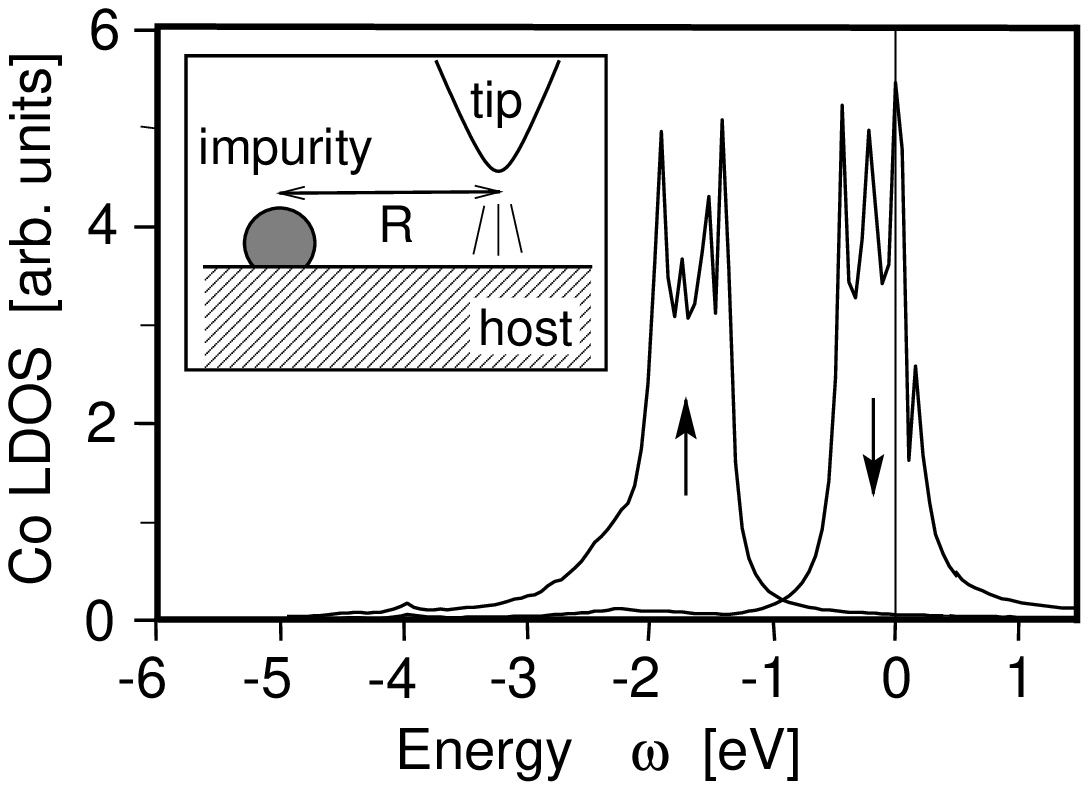}
\hfill
\epsfxsize=5.8cm
\epsfysize=3.5cm
\epsfbox{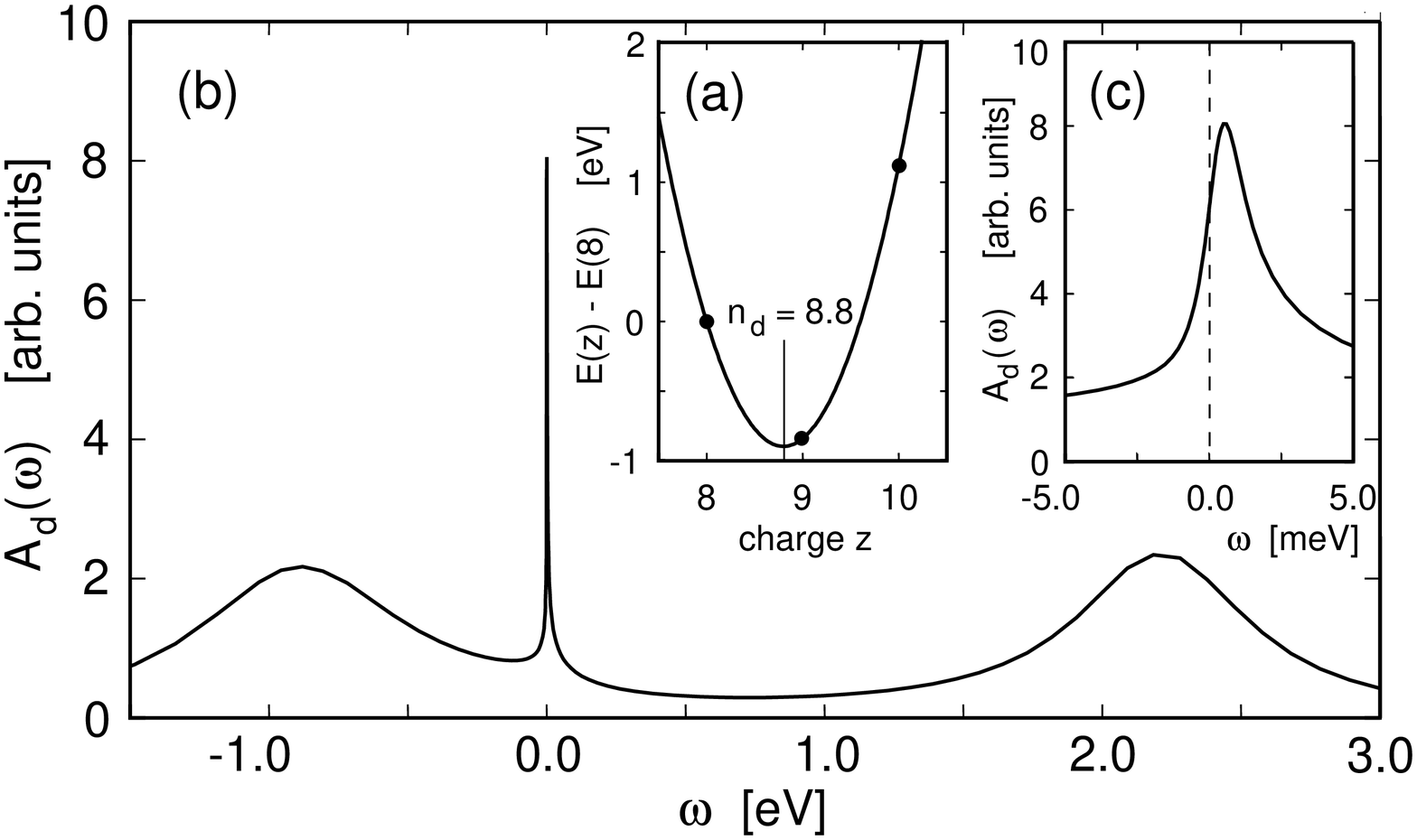}
}
\vspace*{-0.6cm}
\dblcaption
 {\label{fig1} 
    LSDA result for the
    LDOS of a Co impurity on a Au(111) surface, showing
    majority ($\uparrow$) and minority ($\downarrow$) states. The inset
    shows the experimental setup schematically.}
 {\label{fig2} 
    Local d-orbital spectral function 
    $A_d(\omega )$, $T=4$K, showing the broad d resonances 
    near $\overline\varepsilon_d$ and
    $\overline\varepsilon_d+\overline U$ and the shift of the narrow Kondo
    resonance  near $\varepsilon _F$.
    $\bar\varepsilon _d = -0.84$eV, $\bar U = 2.84$eV.}
\end{figure}

%\section*{Introduction}
Due to the Kondo effect, a single magnetic ion in a metallic host 
can produce a narrow resonance at the Fermi level 
in the spectral density of the 
ion's d- (or f-) orbital as well as in the local conduction electron 
density of states (c-LDOS) \cite{Mezei}. However, 
only recently sufficient spatial and energy resolution could be achieved by 
scanning tunneling microscope (STM) spectroscopy to directly measure the 
LDOS correction induced by a single magnetic ion adsorbed on a metal
surface \cite{Li,Madhavan,Manoharan} (see Fig.~\ref{fig1}, inset). 
The electronic density of states, as measured by the $dI/dV$ 
characteristics of the STM, shows a narrow resonance at the Fermi level 
whose asymmetric line shape resembles that of a Fano resonance \cite{Fano}
and changes with the distance $R$ from the ad-atom.

As seen below, the asymmetric Fano line shape 
arises from the interference between electrons traveling in to
%% Double captions:
\inxx{captions,figure}     
the impurity and the outgoing scattered waves which carry information
about the complex, strongly energy-dependent Kondo scattering amplitude.
We emphasize that the Fano line shape is not primarily due to the 
interference between 
electrons tunneling from the STM tip into the conduction band and into 
the d-orbital, respectively. This may be an additional 
effect \cite{Madhavan,Schiller},
which is, however, small, since the d- or f-orbital is deeply localized inside 
the atomic core. The unimportance of this latter interference can be
inferred explicitly from the experiment reported in Ref.~\cite{Manoharan},
where the c-LDOS correction due to a Co ion placed in one focus of 
an elliptic quantum corral was spatially separated from the Co d-orbital by
mapping the c-LDOS into the other focus, 
where still a Fano resonance was observed by STM spectroscopy.       
Therefore, the Fano line shape is an intrinsic property of the c-LDOS
and not an artefact of the STM measurement technique.
In the present article we summarize a detailed  
theoretical description \cite{UKSZZ} of the
Kondo line shape as measured by STM in the vicinity of a single magnetic
ion, taking into account tunneling from the STM into the conduction band only.
For concreteness, the numerical calculations are focused on Co on a 
Au(111) surface.

A systematic STM study of the local electronic structure of individual
transition-metal impurities on Au surfaces was performed by 
Jamneala {\it et al.} \cite{Jamneala}. 
In order to obtain a semi-quantitative understanding of these results 
we have developed a method to combine electronic structure calculations 
with strongly correlated methods to describe the many-body Kondo resonance
\cite{UKSZZ}. The material-specific electronic structure calculations
were performed using the semi-relativistic, screened 
Korringa-Kohn-Rostoker method
\cite{skkr} in combination with the local spin-density approximation
(LSDA) \cite{lsda}, as was done previously \cite{Weissmann}
for similar problems. However, the LSDA can give reliable estimates 
only for quantities which in bulk ferromagnets are correctly described
by a mean field theory. As such quantities we take the 
intra-d-orbital Coulomb repulsion $U$, which is related to the
mean field Stoner splitting, the average d-orbital occupation number $n_d$
and the effective hybridization $\Delta$. From these an effective
single-orbital Anderson impurity model $H^{eff}$ may be constructed 
\cite{UKSZZ} to describe 
the low-energy spin fluctuations of a single magnetic ion which generate
the Kondo effect:
The charge states $z=0,1,\dots ,10$ of the Co ion have energies
$E(z)=z\varepsilon _d +Uz(z-1)/2$ (Fig.\ref{fig2} (a)), 
where $\varepsilon _d $ is the 
single-particle energy of the d-orbital and the system fluctuates
between $z=8,9,10$ only. The ground state is defined by
$dE(z)/dz| _{z=n_d} \equiv 0 = \varepsilon _d +U(n_d-1/2)$. From these relations
the parameters $\bar\varepsilon _d$, $\bar U$ of $H^{eff}$ are
determined via $\bar\varepsilon _d = E(9)-E(8)$ and  
$2\bar\varepsilon _d +\bar U= E(10)-E(8)$ \cite{note}. 
The resulting Co d-orbital spectral function $A_d(\omega )$,
calculated using the Non-Crossing Approximation (NCA) \cite{Bickers},
is shown in Fig.~\ref{fig2} (b) and (c). 
The Kondo temperature was estimated as $T_K\simeq 52$K.

\begin{figure}[ht]
\centerline{
\epsfxsize=5.8cm
\epsfysize=3.2cm
\epsfbox{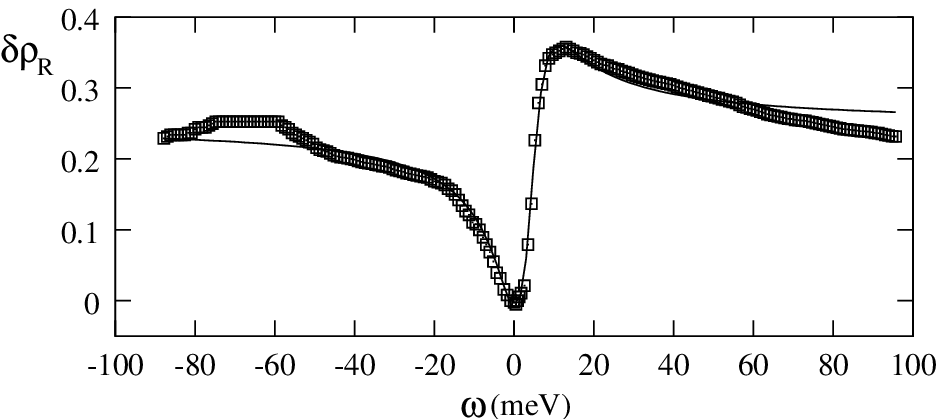}
\hfill
\epsfxsize=5.8cm
\epsfysize=3.2cm
\epsfbox{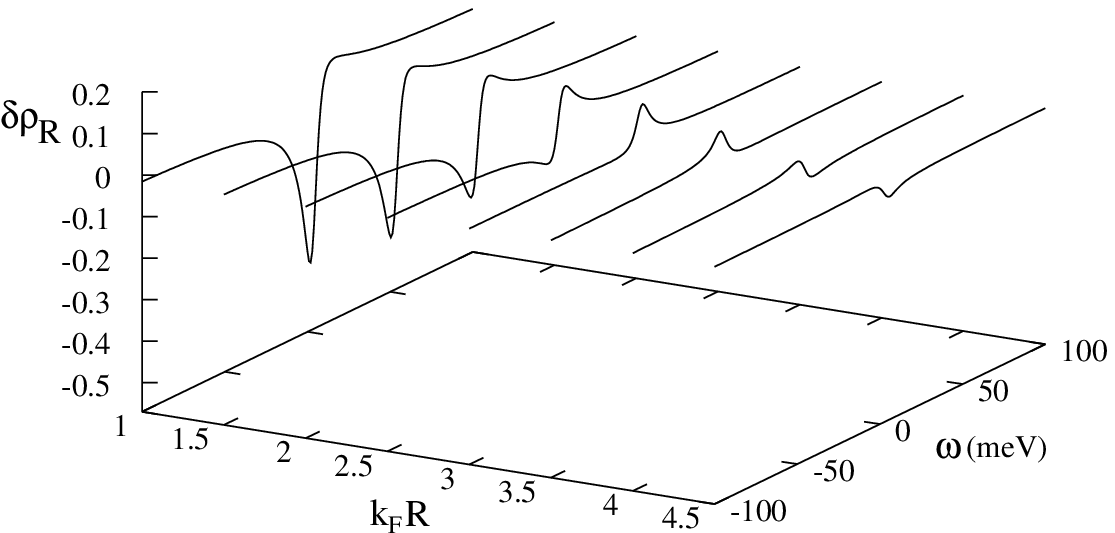}
}
\vspace*{-0.6cm}
\dblcaption
  {\label{fig3} 
      Fit of Eq.~({\ref{roveg}}) (solid line) 
      to the experimental data {\cite{Madhavan}} (squares) at $R=0$.
      The fit parameters are $q_{R=0} =0.66$, $C_{R=0}=0.95$ and
      $\varepsilon_K=3.6$meV,  $T_K=5$meV$\sim 50$K.}
  {\label{fig4} Qualitative dependence of the line shape of the
      tunneling DOS on the distance of the tip from the
      impurity using Eq.~(\ref{roveg}).}
\label{fig4} 
\end{figure}

For the analytical
treatment of the Fano line shape we model $A_d(\omega)=\frac{1}{\pi}{\rm Im}
G_d(\omega-i\delta)$ as a sum of three Lorentzians \cite{UKSZZ} 
(see Fig. \ref{fig2}) and calculate the correction to the conduction electron
Green's function due to the presence of the impurity as
$\delta {\cal G}_{R}(i\omega_n) =
  {\cal G}^{(0)}_{R}(i\omega_n) \,
  t (i\omega_n) \, {\cal G}^{(0)}_{R}(i\omega_n)$,
where ${\cal G}^{(0)}_{R}(i\omega_n)$ is the one-electron
surface Green's function (per spin) of the unperturbed metal 
and, according to the Anderson model, the exact conduction 
electron $t$-matrix is given as $t (i\omega_n)=
\frac{\Delta}{\pi\rho_0}G_{d}(i\omega_n)$.
Thus we obtain for the perturbation in the tunneling c-LDOS
at distance $R$ \cite{UKSZZ}
 \begin{equation}
  \delta\rho_{R}(\omega) =\frac1\pi{\rm Im} \delta {\cal G}_{R}(\omega-i\delta)=
   \frac{[{\rm Im} {\cal G}^{(0)}_{R}(\omega -i\delta)]^2}{\pi\rho _0}\,
  \biggl \{
%  \frac{\Delta Z_K}{\pi T_K}
  \frac{(\varepsilon + q_{R})^2}{\varepsilon^2+1}-1+ C_R \biggr \}
\label{roveg}
\end{equation}
where $\varepsilon=(\omega-\varepsilon_K)/{T_K}$
and $q_{R}={{\rm Re} {\mathcal G}^{(0)}_{R}(\omega-i\delta)}
/{{\rm Im} {\mathcal G}^{(0)}_{R}(\omega-i\delta)}$ were introduced, with
$\varepsilon_K$ the shift of the Kondo resonance w.r.t. the Fermi level
and $\rho_0$ the bare conduction electron DOS.
$C_R$ arises from potential scattering by the
d-level and corresponds to a weakly energy dependent
Friedel oscillation.
The first part of Eq.~(\ref{roveg}) coming
from the scattering by the Kondo resonance gives a Fano line
shape in the tunneling c-LDOS, controlled by the parameter 
$q_{R}= {\rm Re}{\cal G}^{(0)}_{R}(0) /{\rm Im}{\cal G}^{(0)}_{R}(0)$,
in complete analogy to Fano's asymmetry parameter $q$ \cite{Fano}.
The fit of Eq. (\ref{roveg}) to the experimental data for a Co atom on a
Au(111) surface \cite{Madhavan} gave excellent
agreement (Fig.~\ref{fig3}), the fit parameters being consistent with
the  predictions of the NCA calculation in combination with 
LSDA \cite{UKSZZ}.

To calculate the distance dependence of $q_{R}$ and $C_{R}$, i.e. of the
line shape, the tunneling of electrons from the tip (1)
into the 3-dimensional Au bulk states as well as (2) into the 2-dimensional
Au(111) surface band \cite{Chen} was considered \cite{UKSZZ} (Fig. \ref{fig4}). 
Comparing the wave length of the oscillating line shape with the 
known Fermi wave numbers of the Au bulk and the surface band
\cite{Chen} one can conclude that tunneling occurs predominantly into
the Au(111) surface band. The quantitative distance dependence of the line shape
is sensitive to details of the conduction band structure. 
\inxx{captions,figure}

%\begin{acknowledgments}

Discussions with R. Berndt, M. F. Crommie and W.-D.~Schneider are gratefully
acknowledged. J.K. is grateful for the hospitality of the Condensed Matter 
Physics group, TU Budapest. This work was supported by the
Humboldt foundation (A.Z.), by the OTKA Postdoctoral Fellowship D32819 (O.\'U.),
by Hungarian grants
OTKA T024005, T029813, T030240 and by DFG through SFB 195 (J.K.).
%\end{acknowledgments}


\begin{chapthebibliography}{1}
\bibitem{note} As a consistency check, the relation $\bar U = U$ should be 
(and approximately is \cite{UKSZZ}) fulfilled.
\end{chapthebibliography}
\end{document}